\documentclass[]{article}
\usepackage[utf8]{inputenc} 
\usepackage[T1]{fontenc}    
\usepackage{hyperref}       
\usepackage{url}            
\usepackage{booktabs}       
\usepackage{amsfonts}       
\usepackage{microtype}      
\usepackage{xcolor}         
\usepackage{braket}         
\usepackage{amsmath}        
\usepackage{graphicx}       
\usepackage[ruled,vlined]{algorithm2e}  
\usepackage{enumitem}       
\usepackage{amsthm}         
\usepackage{xspace}         
\usepackage{subfig}         
\usepackage{bbold}          
\usepackage{float}          
\usepackage{svg}            
\usepackage{array}

\usepackage{chngcntr}

\newcounter{mysfig}
\counterwithin{mysfig}{figure}

\renewcommand\themysfig{\thefigure(\alph{mysfig})}
\makeatletter
\newcommand\Scaption[1]{%
\refstepcounter{mysfig}%
\vskip.5\abovecaptionskip
  \sbox\@tempboxa{\small\themysfig~#1}%
  \ifdim \wd\@tempboxa >\hsize
    \small\themysfig~#1\par
  \else
    \global \@minipagefalse
    \hb@xt@\hsize{\hfil\box\@tempboxa\hfil}%
  \fi
  \vskip\belowcaptionskip}
\makeatother

\usepackage{todonotes}
\usepackage{comment}

\hypersetup{
    colorlinks=true,
    urlcolor=cyan,
}

\newtheoremstyle{cited}%
  {3pt}
  {3pt}
  {\itshape}
  {}
  {\bfseries}
  {.}
  {.5em}
  {\thmname{#1} \thmnumber{#2} \thmnote{\normalfont#3}}

\theoremstyle{cited}

\theoremstyle{plain}

\theoremstyle{definition}

\newcommand{\R}{\mathbb{R}}

\SetKwComment{Comment}{$\vartriangleright$}{}

\usepackage{authblk}
\usepackage{subfig}
\usepackage{svg}

\newcommand{\rmsd}{RMSD~}

\title{Assessing Cost Hamiltonian Reliability in Quantum Protein Structure Prediction}
\author[1,2]{Mathieu Roget}
\author[3]{Cedric Damour}
\author[1,4]{Frederic Cadet}
\author[2]{Jingbo Wang}
\affil[1]{University Paris City \& University of Reunion, 75015, Paris, France}
\affil[2]{Centre for Quantum Information, Simulation and Algorithms, University of Western Australia, Australia} 
\affil[3]{ENERGYLab, University of Reunion, Saint Denis, France}
\affil[4]{PEACCEL, AI for Biologics, Paris, France} 

\date{}

\begin{document}
	
	\maketitle
	
	\begin{abstract}

        In variational quantum algorithms, QAOA, and quantum annealing, the cost Hamiltonian defines the optimization landscape explored by the quantum hardware; however, in many application-driven formulations, this Hamiltonian is a simplified proxy for the true task-level objective. Using lattice-based quantum protein structure prediction as a case study, we investigate whether the contact-energy cost Hamiltonian commonly used in this setting is sufficiently aligned with structural accuracy, as measured by RMSD against experimentally determined structures. Through this specific problem, we show the importance of studying the reliability of the cost Hamiltonian independently from the quantum approach used. This work shows that, for small peptides and on average, the energy landscape of the considered cost Hamiltonian is not correlated well enough to the actual error to provide meaningful predictions. Moreover, this correlation was estimated through Monte-Carlo sampling for larger instances. It shows an increase of said correlation for larger problem instances and when more interaction shells are considered. This investigation illustrates the meaningfulness of investigating cost Hamiltonian relevance independently from the quantum algorithm used. 

	\end{abstract}
	
	\section{Introduction}

    In quantum optimization, the cost Hamiltonian is the object that connects the computational problem to the quantum algorithm. Its design determines which configurations are energetically favored and therefore what the quantum optimizer is expected to return. In application-driven settings, however, the cost Hamiltonian is often a simplified or hardware-compatible proxy for the actual task-level objective\cite{bennett2025benchmarking, zering2025benchmarking, green2026quantum}. This creates a fundamental validation problem: even if the quantum optimizer identifies low-energy states efficiently, these states may not correspond to high-quality solutions for the original problem unless the encoded Hamiltonian is sufficiently aligned with the target metric.

    Protein structure prediction (PSP) \cite{levinthal1968there, huang2023protein, anfinsen1973principles, berg2007biochemistry, ebrahimi2023engineering} provides a particularly relevant test case for this question. Existing quantum formulations typically reduce the structure to a coarse-grained lattice model and encode contact-based interaction energies into a cost Hamiltonian\cite{dogaPerspectiveProteinStructure2024, liQuantumAlgorithmProtein2025, robertResourceefficientQuantumAlgorithm2021, boulebnanePeptideConformationalSampling2023, zhangQDockBankDatasetLigand2025, perdomo2012finding}. While this formulation makes the problem compatible with VQE\cite{peruzzo2014variational}, QAOA\cite{farhi2014quantum}, quantum annealing\cite{kadowaki1998quantum}, and related approaches\cite{bennett2024non, matwiejew2024quantum, matwiejew2024quantum, matwiejew2023quantum, matwiejew2022quop_mpi, marsh2021framework, marsh2020combinatorial, marsh2019quantum}, it also introduces a gap between the quantity being optimized and the structural accuracy ultimately desired. This gap motivates an independent assessment of cost-Hamiltonian relevance before drawing conclusions about the usefulness of the quantum optimization step.

	Previous works on this approach have researched scalable implementations of quantum schemes such as quantum annealing\cite{perdomo2012finding}, VQE\cite{robertResourceefficientQuantumAlgorithm2021, zhangQDockBankDatasetLigand2025} or QAOA\cite{boulebnanePeptideConformationalSampling2023} for this problem. While several implementations on quantum devices have been realized, and have shown the algorithms remarkable performances for converging toward minimum or low cost solutions, the relevance of said cost function remains to be seen.
	
	This work investigates the feasibility of this quantum approach and points out how the cost function plays a preponderant role in degrading the predictions' quality. In particular, we show that the quantum optimization algorithm might be optimizing the wrong objective function. It is important to develop methodologies to evaluate these objective functions' relevance even for problem instances too large to be run on current hardware. The methodology of this work can be applied beyond the PSP problem. Indeed, it is common, because of the problem's nature or because of a quantum engineering trade-off, for the cost function used by VQE, QAOA or quantum annealing to differ from the actual error\cite{farhi2014quantum, zylberman2025efficient}. In these situations, the reliability of the cost function must be evaluated independently from the quantum scheme used to solve the optimization task.
	
	Section \ref{sec:prelim} introduces in details the quantum approach and how the cost function is calculated. In Section \ref{sec:methods}, we explain the workflow of the numerical experiments of this work. Section \ref{sec:results} presents and discusses the results. The takeaway message is threefold: (i) the cost function studied here does not capture the problem and is not suitable for small peptides; (ii) the cost function relevance to the problem can be estimated even for large proteins through Monte Carlo methods; (iii) by considering higher order interaction terms on proteins of length 75 amino acids and more, the correlation between cost and error can be significantly increased. 
    The improvement observed when additional interaction shells are included highlights a central trade-off in quantum algorithm design for biomolecular optimization. Simpler cost Hamiltonians are easier to encode and may require fewer quantum resources, but they may provide a poor approximation of the structural objective. Conversely, richer Hamiltonians can better reflect the underlying protein-structure problem, but they generally introduce denser interaction graphs, additional terms, higher-order dependencies, deeper circuits, or more complex penalty structures\cite{zylberman2025efficient, rajak2023quantum, lucas2014ising, yarkoni2022quantum}. Quantifying this trade-off is essential for identifying which formulations are both structurally meaningful and realistic for near-term quantum hardware.

	\section{Preliminaries}\label{sec:prelim}
	This section introduces the approach of \cite{robertResourceefficientQuantumAlgorithm2021} and later \cite{dogaPerspectiveProteinStructure2024, liQuantumAlgorithmProtein2025, boulebnanePeptideConformationalSampling2023}. 
	
	Let us note an instance of PSP as the sequence $s = \left(s_i\right)_{1\leq i\leq n} \in \{1,\ldots,20\}^n$. $s_i$ is the $i^{\text{th}}$ amino acid of the protein of interest, with the amino acids numbered from $1$ to $20$. The main idea of this quantum approach is to transform the PSP problem into an integer optimization problem. 
	
	\paragraph{Solution space}
	Solutions to the PSP problem are encoded as self avoiding walks on a tetrahedral lattice. In practice, a solution to a PSP instance $s = \left(s_i\right)_{1\leq i\leq n} \in \{1,\ldots,20\}^n$ of length $n$ as a list $t = \left(t_i\right)_{1\leq i\leq n-1} \in \{1,\ldots,4\}^{n-1}$. This list $t$ encodes $n-1$ turns on the tetrahedral lattice. A protein structure can be calculated from such a list of turns according to the following scheme. First, place $s_1$ at an arbitrary position on the tetrahedral lattice. Then, follow the turns $t_i$ to place $s_{i+1}$ on the lattice. Finally extract the coordinates of each amino acid $s_i$. Those are the coordinates of the $C_\alpha$ atoms. All-atom coordinates can be reconstructed from $C_\alpha$ coordinates via tools such as pulchra \cite{rotkiewicz2008fast}.
	
	\paragraph{Cost function}
	In order to define an optimization problem, a solution space and a cost function are needed. Thus, we are looking for a function $p:\{c_1, \ldots,c_n\} \to \R$ that returns low values for good predictions and high values for bad predictions. In this function, $\{c_1, \ldots,c_n\} \in (\R^3)^n$ is the list of the predicted xyz coordinates of $C_\alpha$ atoms. Here, we use the sum of the contact energies between the residues. More precisely, we define 
	\begin{description}
		\item[$d_\text{max}$] the contact distance between two amino acids,
		\item[$k_\text{min}$] the minimum sequence distance between two amino acids,
		\item[$\left(\epsilon_{i,j}\right)_{1\leq i,j\leq 20}$] the contact energies of pairs of amino acids.
		\item[$\lambda = (\lambda_1,\ldots, \lambda_l)$] the weight coefficient for each order of interaction. 
	\end{description}
	The cost function writes
	$$
	p(c,s) = \sum_{1\leq i<j\leq n} \sum_{m=1}^l \delta_{i,j,m}(c)\lambda_m\epsilon_{s_i,s_j},
	$$
	where
	$$
	\delta_{i,j,m}(c) = \left\{\begin{matrix}
		1 & \text{if } |i-j| > k_\text{min} \text{ and } (m-1)d_\text{max} < \Vert c_i-c_j\Vert_2 \leq m d_\text{max}\\
		0 & \text{otherwise}
	\end{matrix}\right..
	$$
	This cost function essentially sums the contact energies for all amino acids at euclidian distance less than $d_\text{max}$ and sequence distance more than $k_\text{min}$ for the first order interactions. It does the same for the amino acids at euclidian distance between $d_\text{max}$ and $2d_\text{max}$ for second order and so on. The contact energies $\epsilon_{i,j}$ are obtained by analyzing known structures. Usually, the MJ coefficients \cite{miyazawaResidueResiduePotentials1996} are used. A penalization term can be added to this cost function to remove unwanted behaviors. For instance, to penalize solutions with collisions (solutions where two $C_\alpha$ atoms have the same coordinates).

    In the quantum optimization schemes considered in previous works, this classical cost function is the objective that is encoded into the corresponding cost Hamiltonian. The quantum algorithm therefore does not optimize the RMSD or the structural error directly, but the energy landscape induced by this cost function over the feasible conformational space. Assessing the relevance of the cost function is thus a prerequisite for assessing whether the associated cost Hamiltonian provides a meaningful optimization landscape for protein structure prediction.

\begin{figure}
		\includegraphics[width=\textwidth]{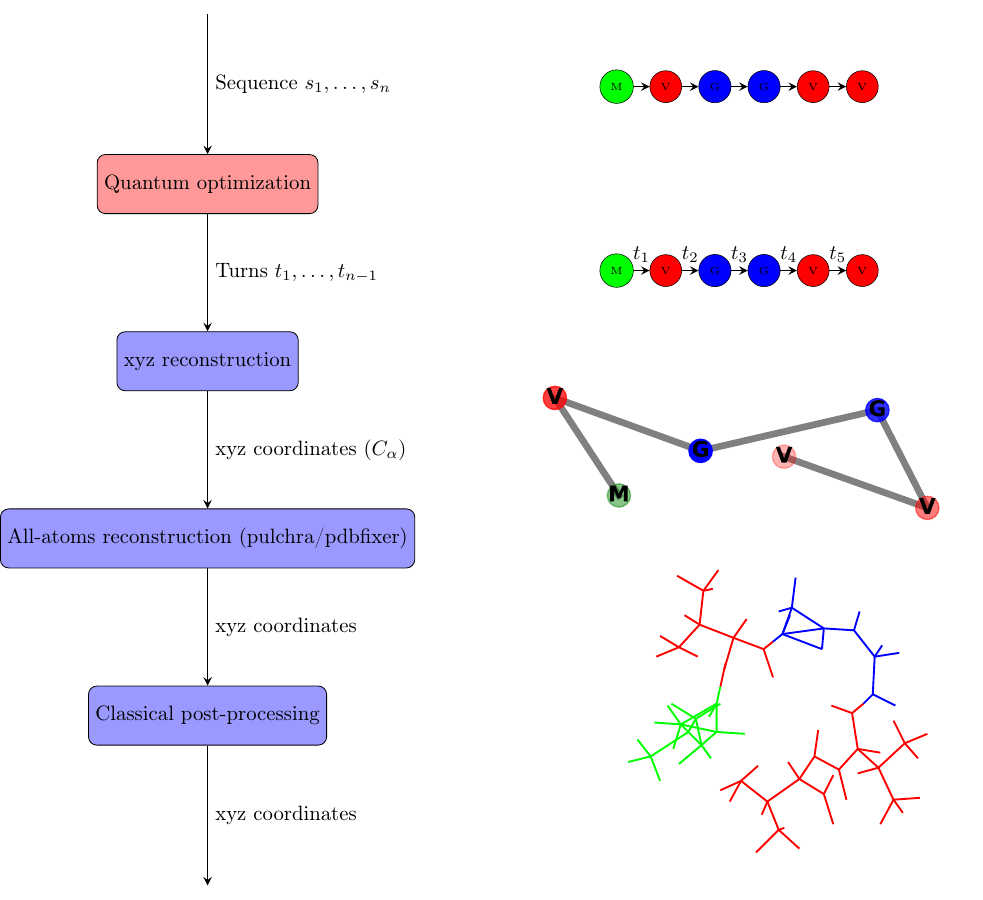}
		\caption{Full quantum scheme for solving the PSP problem (on the left) with examples of the various objects manipulated (on the right).}
		\label{fig:PSPscheme}
	\end{figure}

	\paragraph{Full scheme}
	The full scheme, from a pdb instance to an all-atom xyz prediction includes the following:
	\begin{enumerate}
		\item Formulate the problem using a coarse grain model on a tetrahedral lattice and solve it with a Quantum algorithm like VQE or QAOA.
		\item Convert the list of turns obtained by the quantum algorithm into the xyz coordinates of the $C_\alpha$ atoms.
		\item Reconstruct an all-atom prediction.
		\item An optional classical post-processing of the solution to improve it.
	\end{enumerate}
	This scheme is pictured in Figure \ref{fig:PSPscheme}. 
	
	\paragraph{Related works}
	The first introduction, to our knowledge, of this quantum formalization of the PSP problem for the tetrahedral lattice is by \cite{robertResourceefficientQuantumAlgorithm2021}. They use a tetrahedral lattice and the VQE quantum algorithm. An important point of this work is how they smartly change the cost function. They manage to provide an altered version that admits an efficient quantum implementation without degrading its overall meaning. This is crucial for an efficient implementation on a quantum computer. An attempt to replace the VQE algorithm with QAOA by \cite{boulebnanePeptideConformationalSampling2023} showed a degradation of the predictions. In \cite{liQuantumAlgorithmProtein2025}, the authors have investigated the use of a face centered cubic lattice instead of a tetrahedral lattice. They show that a face centered cubic lattice is able to describe more accurately the structure of a protein and could, in theory, provide better predictions. Several works have also focused on the quality of the prediction using indicators like the \rmsd. This is the case of \cite{dogaPerspectiveProteinStructure2024} which compares this quantum approach to AlphaFold\cite{li2025advantages, jumper2021highly, nassar2021protein, ruff2021alphafold}. It discusses which instances are susceptible to benefit from a quantum approach. \cite{zhangQDockBankDatasetLigand2025} tries this approach on real quantum computers for several small peptides and shows improved performances for some of them compared to AlphaFold. 
    The goal of our work is not to benchmark quantum protein structure prediction against state-of-the-art deep-learning methods for structure prediction. Instead, we focus on a more fundamental prerequisite for quantum optimization approaches: whether the cost Hamiltonian being optimized is sufficiently aligned with structural accuracy. This question is independent of the performance of any specific quantum optimizer and must be addressed before meaningful comparisons with classical or AI-based prediction pipelines can be made.

	\paragraph{Limits}
	This scheme presents two points of interest. First, it allows the use of quantum computers and their speed-up. Second, it provides an energy based approach that could complement AI based methods. However, this scheme also makes several approximations which might degrade the results. This is in addition of the quantum optimizer itself not being guaranteed to find the best overall solution. These approximations are 
	\begin{description}
		\item[1. Coarse grain model] Predicting only the positions of the $C_\alpha$ atoms and reconstructing the all-atom prediction afterward is a rather common approximation.
		\item[2. Discrete solutions on a lattice] Limiting the coordinates to be on a lattice is a strong limitation regarding the angles between amino acids.
		\item[3. Cost function] The cost function is emulating the calculation of the free energy. It is based on the physical principle that the protein folds into its state with minimum free energy. But the cost function we are using calculates the contact energies instead of the free energy and uses knowledge based coefficients.
	\end{description}
	This paper focuses on investigating the performances and relevance of the cost function. It shows that, among the three approximations above, the third one is the most significant. In order to do this, the cost and \rmsd against ground truth (prediction error) are calculated for several feasible solutions.

	\section{Methods}\label{sec:methods}
\begin{figure}
		\centering
		\includegraphics[width=\linewidth]{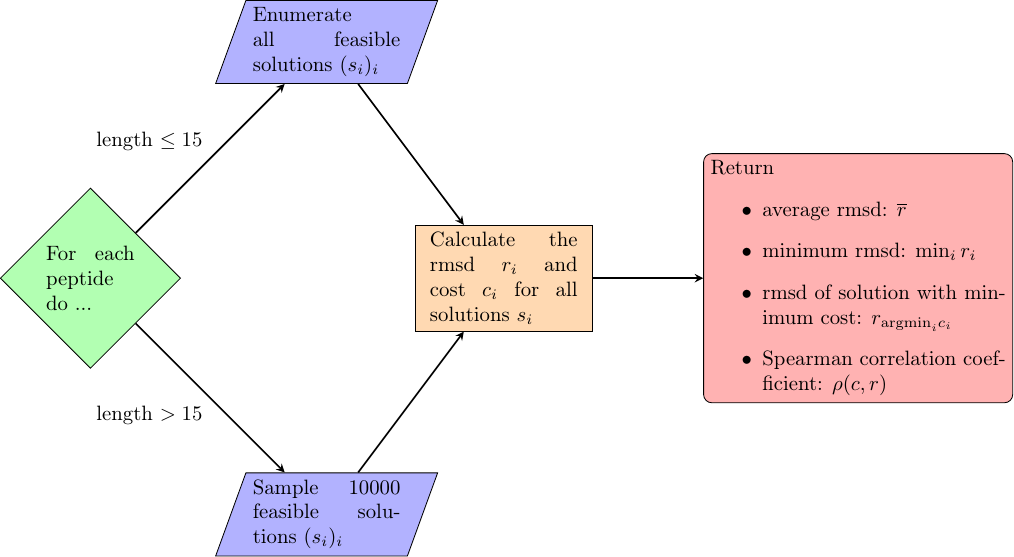}
		\caption{Workflow of the numerical experiments done in this paper.}
		\label{fig:workflow}
	\end{figure}

	This section details the numerical experiments done to evaluate the performance of the scheme introduced in Section \ref{sec:prelim}. A dataset of 12446 peptides (including 4866 unique sequences) of length between 6 and 600 amino acids downloaded from the PDB database has been considered. For each peptide, several feasible solutions are enumerated. For each of them, the cost function and \rmsd (error) against ground truth are computed. Then, the average \rmsd across all solutions, the minimum \rmsd across all solutions and the \rmsd of the solution with the least cost. For each peptide, the Spearman rank correlation between the cost and \rmsd of the solutions is also calculated. Figure \ref{fig:workflow} summarizes the workflow used to obtain the numerical results of this work. For the main benchmark, exact sequence duplicates were removed to avoid over-weighting sequences that appear multiple times in the PDB. Since entries with missing residues or missing $C_\alpha$ atoms had already been discarded, one representative structure per sequence was selected deterministically as the first (pdbid,chain) in lexicographic order. We also repeated the analysis on the full set of filtered PDB instances, including duplicates, and observed the same qualitative trends.
	
	\paragraph{\rmsd}
	The \rmsd designates the Root Mean Square Deviation (\rmsd) against ground truth ($C_\alpha$ coordinates obtained from wet-lab experiments). It is calculated after the two structures have been aligned. It is a standard way of measuring the difference between two structures and is used as a prediction error. The solution with minimum \rmsd is the solution closest from ground truth that can be expressed on the tetrahedral lattice. It measures the degradation of the prediction due to using a tetrahedral lattice. As for the \rmsd of the solution with minimum cost, it is the best result one can hope to obtain by running the quantum optimization.

	\paragraph{Spearman's rank correlation coefficient}
	A well known coefficient to measure the linear correlation between two random variables is the Pearson Correlation Coefficient (PCC). The PCC is a value between $-1$ and $1$. 
	\begin{itemize}
		\item A PCC of $1$ indicates a perfect positive and linear correlation.
		\item A PCC of $-1$ indicates a perfect negative and linear correlation.
		\item A PCC of $0$ indicates no linear correlation.
	\end{itemize}
	In our case, the lower the cost, the most attractive the solution is for our algorithm. At the same time, the lower the \rmsd against ground truth, the better the solution is regarding the PSP problem. So we want a correlation as close to $1$ as possible. However, a good cost function (i.e. relevant to the problem) is not necessarily linearly correlated. It must simply order the solutions the same way the \rmsd (measure of the solution's quality) does. Therefore, instead of calculating the PCC between $\left(c_i\right)_i$ and $\left(r_i\right)_i$, we calculate the PCC between the ranks of these observations. This coefficient $\rho(c,r)$ is called the Spearman's rank correlation coefficient and measure the monotonic correlation between two random variables. In this paper, the correlation coefficient implicitly means the  Spearman's rank correlation coefficient between cost and \rmsd. The mathematical formulas for both coefficients are stated below.
	$$
	\text{PCC}(c,r) = \frac{\sum_{i=1}^n}{(c_i - \overline c)(r_i - \overline r)}{\sqrt{\sum_{i=1}^n}{(c_i - \overline c)^2}\sqrt{\sum_{i=1}^n}{(r_i - \overline r)^2}}, 
	$$
	and 
	$$
	\rho(c,r) = \text{PCC}(\text{rank}(c), \text{rank}(r)).
	$$

	\paragraph{Solutions' enumeration}
	For peptide with a length less or equal to $15$ amino acids, all feasible solutions are enumerated. However, since the number of feasible solutions grows exponentially with the protein's length, we only sample 10000 solutions for larger proteins (of length strictly greater than $15$). The sampling is done using the Pivot algorithm. This method does not allow obtaining minima or maxima across the whole feasible solution space. However, accurate estimations of the Spearman correlation coefficient can still be derived. An important practical implication of this approach is that cost-Hamiltonian relevance can be assessed before implementing the corresponding optimization problem on quantum hardware. By estimating the rank correlation between the encoded cost and the task-level error over sampled feasible configurations, one can identify objective functions that are unlikely to produce meaningful solutions even under ideal optimization. This provides a low-cost validation step for quantum algorithm design and can help prioritize Hamiltonian formulations before committing quantum resources. The 95\% confidence interval's amplitude when using 10000 samples to estimate the Spearman correlation coefficient is approximately of $0.05$. This is sufficiently accurate to support the conclusion of this work. More information on how this value was obtained is available in Appendix \ref{sec:mc}.

	\paragraph{Availability of code and data}
	The numerical results produced for this work and the code used to produce these results are all available on github\footnote{\href{https://github.com/mroget/Paper-Code---Assessing-Cost-Hamiltonian-Reliability-in-Quantum-Protein-Structure-Prediction}{https://github.com/mroget/Paper-Code---Assessing-Cost-Hamiltonian-Reliability-in-Quantum-Protein-Structure-Prediction}}.

	\subsection*{Example for a given peptide}
\begin{figure}
		\begin{center}
			\centering
			
			\subfloat[The solution space: each point is a feasible solution. The three \rmsd values extracted in the workflow of Figure \ref{fig:workflow} are displayed in the legend] {\includegraphics[width=\textwidth]{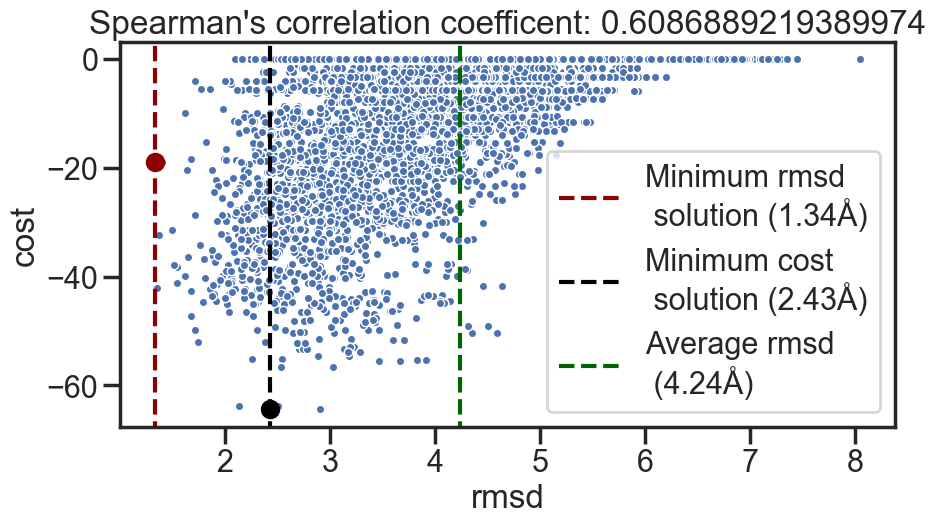}\label{fig:peptide-example:space}}
			
			\subfloat[The solution with minimum \rmsd.]{\includegraphics[width=0.48\textwidth]{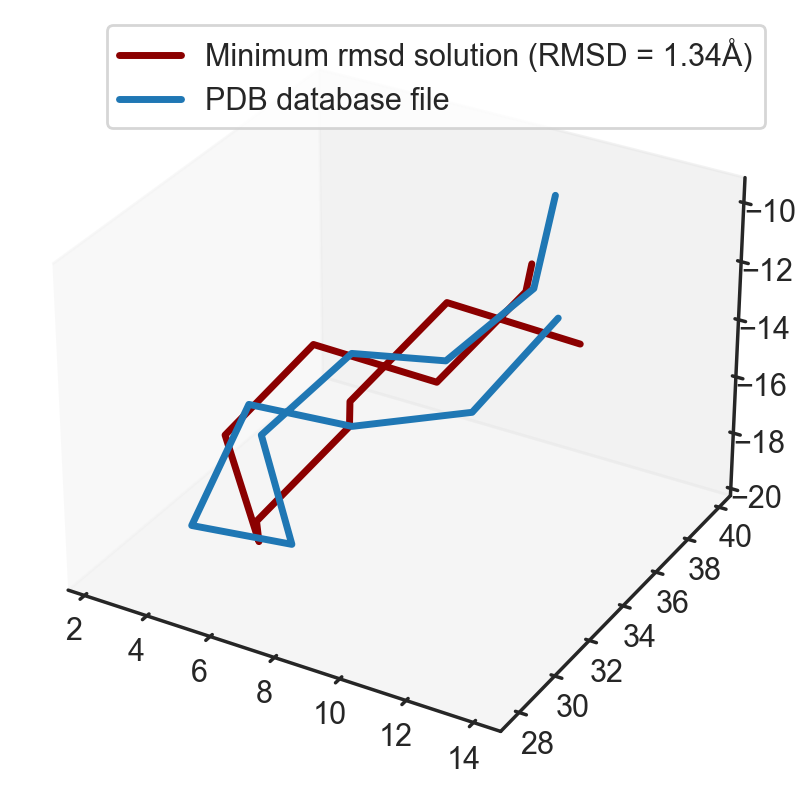}\label{fig:peptide-example:rmsd}}
			\subfloat[The solution with minimum cost.]{\includegraphics[width=0.48\textwidth]{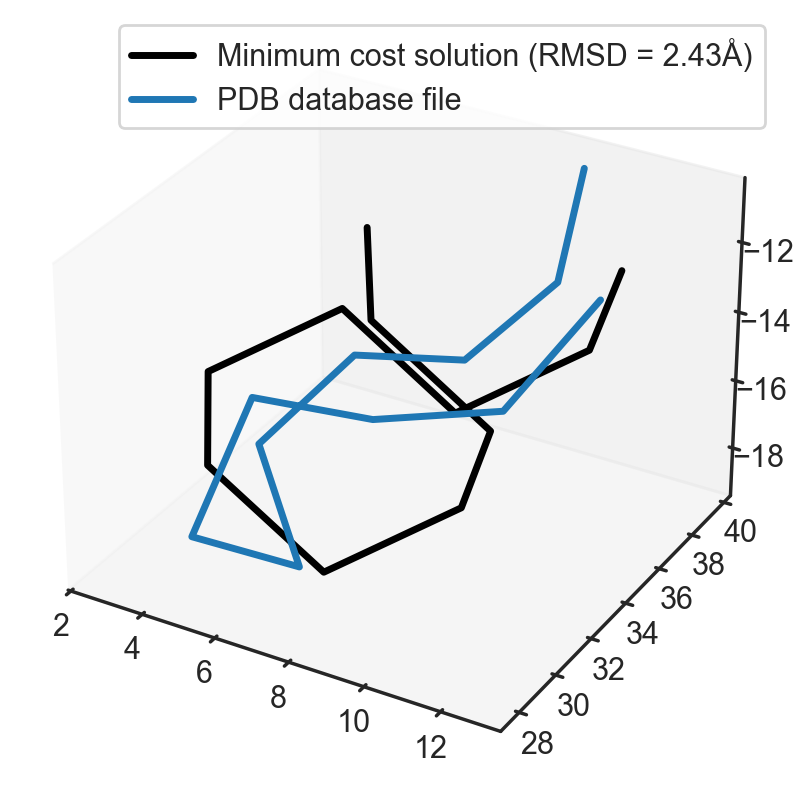}\label{fig:peptide-example:cost}}\\ 
			
			\caption{Solution space for peptide 1I93 of length 11 on the tetrahedral lattice.}
			\label{fig:peptide-example}
		\end{center}
	\end{figure}

	This subsection applies the protocol of Figure \ref{fig:workflow} for peptide of PDB id 1I93 of length 11. Specifically, all Self Avoiding Walks (SAWs) on the tetrahedral lattice with arc length $3.8$ Angstroms are considered. Each of them is converted into a potential solution to the PSP problem. This allows the enumeration of all feasible solutions. For each of them, the cost function and \rmsd (error) against ground truth are computed. Figure \ref{fig:peptide-example:space} displays each solution as a point. The average \rmsd across all solutions is displayed as well. We can see in this example that the solution with minimum \rmsd, which is the actual best prediction one can make on the tetrahedral lattice, is different from the solution with minimum cost. The solution with minimum \rmsd and the solution with minimum cost are displayed in Figures \ref{fig:peptide-example:rmsd} and \ref{fig:peptide-example:cost} respectively.

    \subsection*{Scope and limitations}
    The simulations made for this work are limited to the specific problem of Protein Structure Prediction (PSP) formulated as a combinatorial optimization problem on a tetrahedral lattice. Only a specific class of cost functions (contact based energy), which appears in previous works, is considered. Nevertheless, the approach used to estimate the reliability of the cost function can be generalized to other problems.

	\section{Results}\label{sec:results}
	This section presents and analyses the results obtained through the protocol displayed in section \ref{sec:methods}. We consider two distinct cases: (i) the \rmsd values extracted for peptides of length $\leq 15$ amino acids; (ii) the correlation coefficients for proteins and peptides of length $\geq 13$.
	\rmsd values are not discussed for proteins with a length $>15$ because calculating minimum or maximum across the whole solution space does not scale. The correlation coefficient is not discussed for peptides of length $<13$ because the solution spaces of these peptides contains too many solutions with the same cost.
	
	\subsection{\rmsd values for peptides}
	This subsection presents the numerical results for the cost function relevance on all peptides of length $\leq 15$. Parameters for the cost function are $d_\text{max}=7.8$, $k_\text{min} = 5$ and $\lambda = \left(1\right)$. 
	
\begin{figure}
		\begin{center}
			\centering
			\includesvg[width=\textwidth]{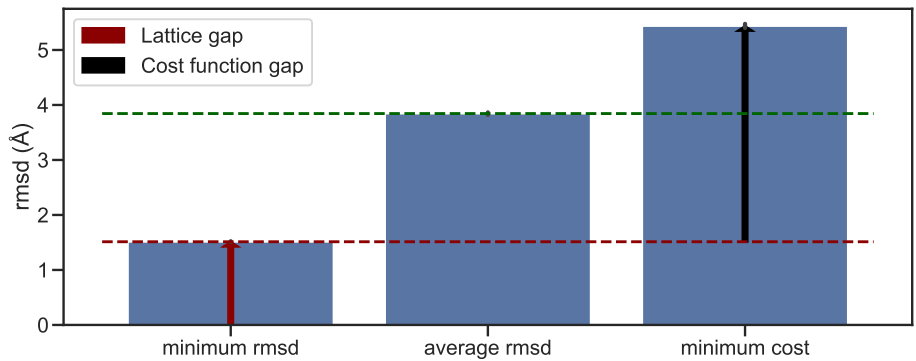}
			\caption{\rmsd values for peptide of length $\leq 15$. Parameters for the cost function are $d_\text{max}=7.8$, $k_\text{min} = 5$ and $\lambda = \left(1\right)$. }
			\label{fig:peptide-results}
		\end{center}
	\end{figure}

	Figure \ref{fig:peptide-results} shows the average values across all peptides of the three metrics: minimum \rmsd, average \rmsd and \rmsd of solution with minimum cost. 
	  The minimum \rmsd quantifies the degradation in accuracy due to using a tetrahedral lattice. This degradation is not negligible at around 1.5 Å. However, we see that the \rmsd of the solution with minimum cost is much higher at around 5 Å. This suggests a significant degradation of the accuracy due to the cost function. Even worse, the average \rmsd across all solutions is, on average across all peptides, better than the optimal solution according to the cost Hamiltonian. This result does not indicate a failure of quantum optimization algorithms per se. Rather, it shows that, for the cost function studied here, the ideal solution of the encoded optimization problem may be structurally inaccurate. The bottleneck is therefore not necessarily the quantum optimizer, but the objective function encoded into the corresponding cost Hamiltonian.

	\subsection{Correlation coefficient for proteins}
	This subsection presents the results for the rank correlation on all peptides and proteins of length $\geq 13$. Parameters for the cost function are $d_\text{max}=7.8$ and $k_\text{min} = 5$. Several $\lambda$ parameters are considered.
	
\begin{figure}
		\begin{center}
			\centering
			\includesvg[width=\textwidth]{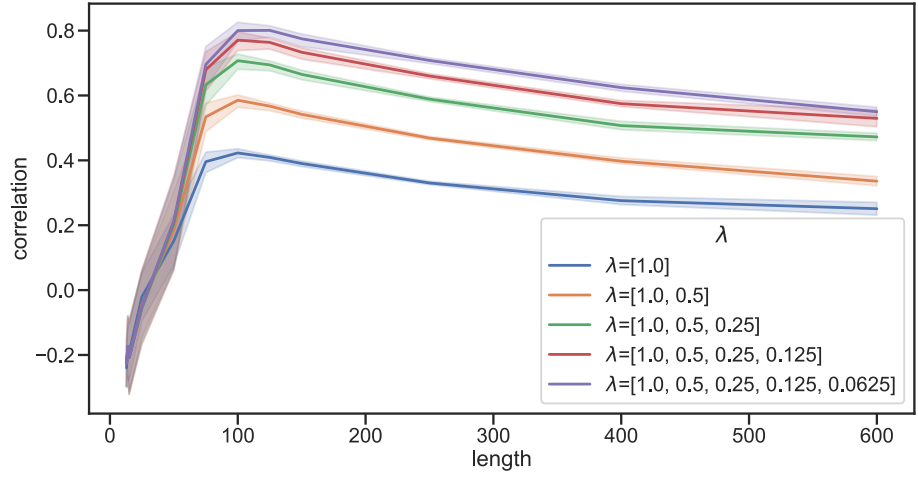}
			\caption{Correlation as a function of the length for proteins for all instances with length $\geq 13$. The higher, the better. Parameters for the cost function are $d_\text{max}=7.8$ and $k_\text{min} = 5$. Several $\lambda$ parameters are considered.}
			\label{fig:protein-correalation-results}
		\end{center}
	\end{figure}

	Figure \ref{fig:protein-correalation-results} displays the average correlation (across all proteins) as a function of the sequence's length. One can see that the correlation is negative for small peptides, null for small proteins of length 50 and becomes positive only for larger proteins. This is not surprising since the MJ \cite{miyazawaResidueResiduePotentials1996} energy coefficients used in the cost function where calculated using large protein (of length higher than 50). It is also hypothesized that a sufficient number of interacting entities must be present for the cost function to be complex enough and capture the problem. Adding orders of interactions, with a $\lambda$ parameter of higher dimension, improves the correlation significantly for larger proteins.

    The improvement observed when additional interaction shells are included highlights a central trade-off in quantum algorithm design for bio-molecular optimization. Simpler cost Hamiltonians are easier to encode and may require fewer quantum resources, but they may provide a poor approximation of the structural objective. Conversely, richer Hamiltonians can better reflect the underlying protein-structure problem, but they generally introduce denser interaction graphs, additional terms, higher-order dependencies, deeper circuits, or more complex penalty structures. Quantifying this trade-off is essential for identifying formulations that are both structurally meaningful and realistic for near-term quantum hardware.

    \subsection{Other cost functions}
    As a sensitivity analysis, we replaced the MJ contact energies by a simpler HP-type contact model \cite{berger1998protein, yue1995test, chan1998protein}. The resulting correlations were comparable to those obtained with the MJ-based cost function, and displayed the same qualitative increase with higher-order interaction terms. This observation should not be interpreted as a general statement about the irrelevance of knowledge-based potentials for protein folding. Rather, it indicates that, in the present $C_\alpha$ lattice-based quantum PSP formulation, the coarse structure of the cost function and the inclusion of additional interaction shells may be at least as important as the precise choice of pairwise contact coefficients. 

    The comparison between MJ and HP-type contact energies further supports the need for systematic cost-Hamiltonian benchmarking. In the present lattice-based formulation, replacing knowledge-based MJ coefficients with a simpler HP-type model leads to comparable qualitative trends. This does not imply that detailed knowledge-based potentials are generally irrelevant for protein folding. Rather, it suggests that, once the problem is projected onto a coarse $C_\alpha$ lattice representation, the structure of the encoded objective — including contact definitions, interaction shells, and sequence-separation constraints — may dominate over the fine details of the pairwise coefficient matrix.

	\subsection*{Remark}
	It is important to note that the most commonly used metric to measure the error of a prediction is the \rmsd. However, this quantity tends to increase with the protein's size. An \rmsd of $3$Å is considered quite accurate for a protein of $100$ amino acids but is awful for a peptide of $6$ amino acids. On the other hand, quantum computing experiments tend to drastically downscale the problem (by using very small instances) and choosing specific instances used as proof of concept. While this is of course essential, it should not stop us from developing relevant cost functions for the average instance that is targeted (i.e. an instance too big to be solved by brute-force).

	\section{Conclusion}
	In this paper, the performance of the quantum scheme for PSP is investigated. This scheme uses quantum optimization such as VQE to find the protein structure with minimal cost. We show that this scheme shows poor accuracy on average largely due to its cost function which approximates the free energy. Indeed, the energy calculation method, simplified to be encoded in a quantum computer, does not correlate with the quality of the solution. To show this, we first calculated the error (\rmsd) of the solution with minimal cost for small peptides. On average, the error of the structure with minimal cost is larger than the error of a structure taken at random. Thus, for short peptides, the minimum-cost conformation has, on average, a larger RMSD than a randomly selected feasible conformation. This result does not indicate a failure of quantum optimization algorithms per se. Rather, it shows that, for the cost Hamiltonian studied here, the ideal solution of the encoded optimization problem may be structurally inaccurate. The bottleneck is therefore not necessarily the quantum optimizer, but the objective function encoded into the Hamiltonian.

	Then we calculated the correlation between the cost and the error. On average, this correlation is negative for small peptides. This again shows that this energy calculation method does not act as a proper cost function. This correlation was then estimated for large proteins through random sampling. Numerical results show that the correlation is higher for larger proteins and increases when considering higher order interactions. This suggests the existence of a trade-off between the implementation complexity and relevance of the cost function. A complex cost function is harder to implement and may use finer models (full atoms or heavy atoms models), but at the same time it captures better the difficulty of the prediction task. This creates a quantum engineering trade-off: richer cost functions may better reflect protein physics, but they introduce higher-order interactions, denser Hamiltonians, additional qubits, or penalty terms that may be difficult to implement on near-term quantum hardware.

    Taken together, these results show that the relevance of the cost function underlying the cost Hamiltonian is a central prerequisite for quantum protein structure prediction. Optimizer performance, circuit design, and hardware implementation cannot be interpreted independently of the objective encoded into the quantum optimization problem. For the contact-energy objective studied here, which defines the corresponding cost Hamiltonian in a quantum implementation, the encoded objective is poorly aligned with structural accuracy for short peptides, while its relevance improves for longer proteins and with additional interaction shells. This suggests that quantum protein structure prediction should be approached not only as an optimization problem, but also as a cost-function and cost-Hamiltonian design problem: the goal is to identify objectives that are simultaneously structurally meaningful and compatible with realistic quantum resources.

    \paragraph{Acknowledgments}
    This work is supported by the European Union (EU) and Region Reunion (FEDER-FSE).

    \paragraph{Author contributions}
    MR developed the algorithms, carried out the experiments, analyzed the results, and drafted the manuscript. CD, JW and FC proposed numerical experiments, contributed to the analysis, and oversaw the project. All authors contributed to and reviewed the manuscript.

	\bibliographystyle{plain}
	\bibliography{biblio}

\begin{thebibliography}{10}

\bibitem{anfinsen1973principles}
Christian~B Anfinsen.
\newblock Principles that govern the folding of protein chains.
\newblock {\em Science}, 181(4096):223--230, 1973.

\bibitem{bennett2024non}
Tavis Bennett, Lyle Noakes, and Jingbo Wang.
\newblock Non-variational quantum combinatorial optimisation.
\newblock In {\em 2024 IEEE International Conference on Quantum Computing and Engineering (QCE)}, volume~1, pages 31--41. IEEE, 2024.

\bibitem{bennett2025benchmarking}
Tavis Bennett, Aidan Smith, Edric Matwiejew, and Jingbo Wang.
\newblock Benchmarking quantum heuristics: Non-variational qwoa for weighted maxcut.
\newblock {\em arXiv preprint arXiv:2505.24191}, 2025.

\bibitem{berg2007biochemistry}
Jeremy~M Berg, John~L Tymoczko, and Lubert Stryer.
\newblock {\em Biochemistry (loose-leaf)}.
\newblock Macmillan, 2007.

\bibitem{berger1998protein}
Bonnie Berger and Tom Leighton.
\newblock Protein folding in the hydrophobic-hydrophilic (hp) is np-complete.
\newblock In {\em Proceedings of the second annual international conference on Computational molecular biology}, pages 30--39, 1998.

\bibitem{boulebnanePeptideConformationalSampling2023}
Sami Boulebnane, Xavier Lucas, Agnes Meyder, Stanislaw Adaszewski, and Ashley Montanaro.
\newblock Peptide conformational sampling using the {{Quantum Approximate Optimization Algorithm}}.
\newblock {\em npj Quantum Information}, 9(1):70, July 2023.

\bibitem{chan1998protein}
Hue~Sun Chan and Ken~A Dill.
\newblock Protein folding in the landscape perspective: Chevron plots and non-arrhenius kinetics.
\newblock {\em Proteins: Structure, Function, and Bioinformatics}, 30(1):2--33, 1998.

\bibitem{dogaPerspectiveProteinStructure2024}
Hakan Doga, Bryan Raubenolt, Fabio Cumbo, Jayadev Joshi, Frank~P. DiFilippo, Jun Qin, Daniel Blankenberg, and Omar Shehab.
\newblock A {{Perspective}} on {{Protein Structure Prediction Using Quantum Computers}}.
\newblock {\em Journal of Chemical Theory and Computation}, 20(9):3359--3378, May 2024.

\bibitem{ebrahimi2023engineering}
Sasha~B Ebrahimi and Devleena Samanta.
\newblock Engineering protein-based therapeutics through structural and chemical design.
\newblock {\em Nature communications}, 14(1):2411, 2023.

\bibitem{farhi2014quantum}
Edward Farhi, Jeffrey Goldstone, and Sam Gutmann.
\newblock A quantum approximate optimization algorithm.
\newblock {\em arXiv preprint arXiv:1411.4028}, 2014.

\bibitem{green2026quantum}
Josh Green and Jingbo Wang.
\newblock Quantum encoding of functions and images with matrix product states.
\newblock {\em Physical Review A}, 113(5):052616, 2026.

\bibitem{huang2023protein}
Bin Huang, Lupeng Kong, Chao Wang, Fusong Ju, Qi~Zhang, Jianwei Zhu, Tiansu Gong, Haicang Zhang, Chungong Yu, Wei-Mou Zheng, et~al.
\newblock Protein structure prediction: challenges, advances, and the shift of research paradigms.
\newblock {\em Genomics, proteomics \& bioinformatics}, 21(5):913--925, 2023.

\bibitem{jumper2021highly}
John Jumper, Richard Evans, Alexander Pritzel, Tim Green, Michael Figurnov, Olaf Ronneberger, Kathryn Tunyasuvunakool, Russ Bates, Augustin {\v{Z}}{\'\i}dek, Anna Potapenko, et~al.
\newblock Highly accurate protein structure prediction with alphafold.
\newblock {\em nature}, 596(7873):583--589, 2021.

\bibitem{kadowaki1998quantum}
Tadashi Kadowaki and Hidetoshi Nishimori.
\newblock Quantum annealing in the transverse ising model.
\newblock {\em Physical Review E}, 58(5):5355, 1998.

\bibitem{levinthal1968there}
Cyrus Levinthal.
\newblock Are there pathways for protein folding?
\newblock {\em Journal de chimie physique}, 65:44--45, 1968.

\bibitem{li2025advantages}
Mandarina Qing~Cheng Li, Sihan Wang, Shi-Ruei Lin, Li~Eric~Ngok Ting, Zhi-Hong Wan, Guodong Xie, and Jane Zhang.
\newblock Advantages and limitations of alphafold in structural biology: Insights from recent studies.
\newblock {\em The Protein Journal}, pages 1--17, 2025.

\bibitem{liQuantumAlgorithmProtein2025}
Rui-Hao Li, Hakan Doga, Bryan Raubenolt, Sarah Mostame, Nicholas DiSanto, Fabio Cumbo, Jayadev Joshi, Hanna Linn, Maeve Gaffney, Alexander Holden, Vinooth Kulkarni, Vipin Chaudhary, Kenneth M.~Merz Jr, Abdullah~Ash Saki, Tomas Radivoyevitch, Frank DiFilippo, Jun Qin, Omar Shehab, and Daniel Blankenberg.
\newblock Quantum {{Algorithm}} for {{Protein Structure Prediction Using}} the {{Face-Centered Cubic Lattice}}, July 2025.

\bibitem{lucas2014ising}
Andrew Lucas.
\newblock Ising formulations of many np problems.
\newblock {\em Frontiers in physics}, 2:74887, 2014.

\bibitem{marsh2019quantum}
S~Marsh and JB~Wang.
\newblock A quantum walk-assisted approximate algorithm for bounded np optimisation problems.
\newblock {\em Quantum Information Processing}, 18(3):1--18, 2019.

\bibitem{marsh2021framework}
S~Marsh and JB~Wang.
\newblock A framework for optimal quantum spatial search using alternating phase-walks.
\newblock {\em Quantum Science \& Technology}, 6(4):045029, 2021.

\bibitem{marsh2020combinatorial}
Samuel Marsh and Jingbo Wang.
\newblock Combinatorial optimization via highly efficient quantum walks.
\newblock {\em Physical Review Research}, 2(2):023302, 2020.

\bibitem{matwiejew2023quantum}
Edric Matwiejew, Jason Pye, and Jingbo~B Wang.
\newblock Quantum optimisation for continuous multivariable functions by a structured search.
\newblock {\em Quantum Science and Technology}, 8(4):045013, 2023.

\bibitem{matwiejew2022quop_mpi}
Edric Matwiejew and Jingbo~B Wang.
\newblock Quop\_mpi: A framework for parallel simulation of quantum variational algorithms.
\newblock {\em Journal of Computational Science}, 62:101711, 2022.

\bibitem{matwiejew2024quantum}
Edric Matwiejew and Jingbo~B Wang.
\newblock Quantum walk informed variational algorithm design.
\newblock {\em arXiv preprint arXiv:2406.11620}, 2024.

\bibitem{miyazawaResidueResiduePotentials1996}
Sanzo Miyazawa and Robert~L. Jernigan.
\newblock Residue -- {{Residue Potentials}} with a {{Favorable Contact Pair Term}} and an {{Unfavorable High Packing Density Term}}, for {{Simulation}} and {{Threading}}.
\newblock {\em Journal of Molecular Biology}, 256(3):623--644, March 1996.

\bibitem{nassar2021protein}
Roy Nassar, Gregory~L Dignon, Rostam~M Razban, and Ken~A Dill.
\newblock The protein folding problem: The role of theory.
\newblock {\em Journal of molecular biology}, 433(20):167126, 2021.

\bibitem{perdomo2012finding}
Alejandro Perdomo-Ortiz, Neil Dickson, Marshall Drew-Brook, Geordie Rose, and Al{\'a}n Aspuru-Guzik.
\newblock Finding low-energy conformations of lattice protein models by quantum annealing.
\newblock {\em Scientific reports}, 2(1):571, 2012.

\bibitem{peruzzo2014variational}
Alberto Peruzzo, Jarrod McClean, Peter Shadbolt, Man-Hong Yung, Xiao-Qi Zhou, Peter~J Love, Al{\'a}n Aspuru-Guzik, and Jeremy~L O’brien.
\newblock A variational eigenvalue solver on a photonic quantum processor.
\newblock {\em Nature communications}, 5(1):4213, 2014.

\bibitem{rajak2023quantum}
Atanu Rajak, Sei Suzuki, Amit Dutta, and Bikas~K Chakrabarti.
\newblock Quantum annealing: An overview.
\newblock {\em Philosophical Transactions of the Royal Society A}, 381(2241):20210417, 2023.

\bibitem{robertResourceefficientQuantumAlgorithm2021}
Anton Robert, Panagiotis~Kl. Barkoutsos, Stefan Woerner, and Ivano Tavernelli.
\newblock Resource-efficient quantum algorithm for protein folding.
\newblock {\em npj Quantum Information}, 7(1):38, February 2021.

\bibitem{rotkiewicz2008fast}
Piotr Rotkiewicz and Jeffrey Skolnick.
\newblock Fast procedure for reconstruction of full-atom protein models from reduced representations.
\newblock {\em Journal of computational chemistry}, 29(9):1460--1465, 2008.

\bibitem{ruff2021alphafold}
Kiersten~M Ruff and Rohit~V Pappu.
\newblock Alphafold and implications for intrinsically disordered proteins.
\newblock {\em Journal of molecular biology}, 433(20):167208, 2021.

\bibitem{yarkoni2022quantum}
Sheir Yarkoni, Elena Raponi, Thomas B{\"a}ck, and Sebastian Schmitt.
\newblock Quantum annealing for industry applications: Introduction and review.
\newblock {\em Reports on Progress in Physics}, 85(10):104001, 2022.

\bibitem{yue1995test}
Kaizhi Yue, Klaus~M Fiebig, Paul~D Thomas, Hue~Sun Chan, Eugene~I Shakhnovich, and Ken~A Dill.
\newblock A test of lattice protein folding algorithms.
\newblock {\em Proceedings of the National Academy of Sciences}, 92(1):325--329, 1995.

\bibitem{zering2025benchmarking}
Matthaus Zering, Jolyon Joyce, Tal Gurfinkel, and Jingbo Wang.
\newblock Benchmarking lie-algebraic pretraining and non-variational qwoa for the maxcut problem.
\newblock {\em arXiv e-prints}, pages arXiv--2512, 2025.

\bibitem{zhangQDockBankDatasetLigand2025}
Yuqi Zhang, Yuxin Yang, Cheng-Chang Lu, Weiwen Jiang, Feixiong Cheng, Bo~Fang, and Qiang Guan.
\newblock Qdockbank: A dataset for ligand docking on protein fragments predicted on utility-level quantum computers.
\newblock In {\em Proceedings of the International Conference for High Performance Computing, Networking, Storage and Analysis}, pages 746--761, 2025.

\bibitem{zylberman2025efficient}
Julien Zylberman, Ugo Nzongani, Andrea Simonetto, and Fabrice Debbasch.
\newblock Efficient quantum circuits for non-unitary and unitary diagonal operators with space-time-accuracy trade-offs.
\newblock {\em ACM Transactions on Quantum Computing}, 6(2):1--43, 2025.

\end{thebibliography}

    \newpage
    \appendix

    \section{Monte Carlo validation}\label{sec:mc}
    This section presents the validation test for the Monte-Carlo estimation of the Spearman correlation coefficient.

    For several proteins of various length, the following scheme was followed:
    \begin{enumerate}
        \item 100 independent estimations of the Spearman correlation coefficient $(\rho_1, \ldots, \rho_{100}$ are calculated.
        \item The $2.5\%$ quantile $q_{2.5}$ and the $97.5\%$ quantile $q_{97.5}$ of the list $(\rho_i)_i$ are calculated.
        \item The amplitude of the $95\%$ confidence interval $\epsilon = q_{97.5} - q_{2.5}$ is calculated and displayed.
    \end{enumerate}

\begin{figure}[H]
        \centering
        \includegraphics[width=0.8\linewidth]{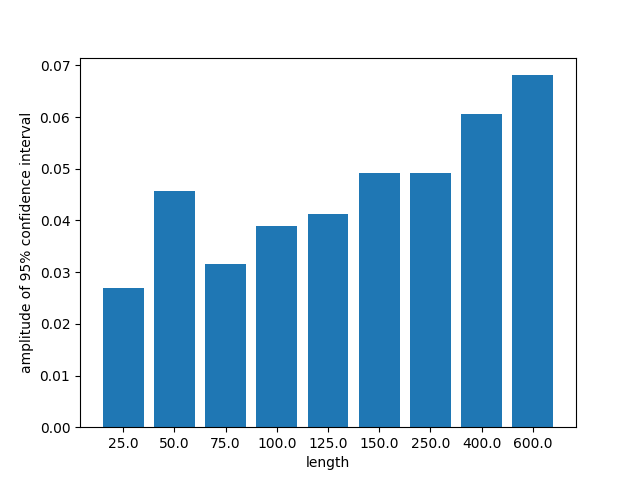}
        \caption{Monte-Carlo validation for instances 7CLVC, 1FDMA, 8BWFE, 6MRDU, 7EHWA, 7MEXB, 5L66O, 7V0NB and 7RWAA (first four letters are the PDB identifier, last letter is the chain identifier).}
        \label{fig:mc}
    \end{figure}

    The results for several proteins, shows a $95\%$ confidence interval amplitude of approximately $0.05$. Figure \ref{fig:mc} shows the amplitude in function of the protein's length. The code to reproduce this experiment is available in the github repository\footnote{\href{https://github.com/mroget/Paper-Code---Assessing-Cost-Hamiltonian-Reliability-in-Quantum-Protein-Structure-Prediction}{https://github.com/mroget/Paper-Code---Assessing-Cost-Hamiltonian-Reliability-in-Quantum-Protein-Structure-Prediction}}.

\end{document}